\documentclass[twocolumn,english,prl, showpacs]{revtex4-1}
\usepackage[T1]{fontenc}
\usepackage[latin9]{inputenc}
\usepackage{mathtools}
\usepackage{amsmath}
\usepackage{amssymb}
\usepackage{cancel}
\usepackage{wasysym}
\usepackage{mathdots}
\usepackage{stmaryrd}
\usepackage{stackrel}
\usepackage{graphicx}
\PassOptionsToPackage{version=3}{mhchem}
\usepackage{mhchem}
\usepackage{esint}
\usepackage{soul}
\usepackage{epstopdf} % CJBF added
\usepackage{color} % cjbf

\makeatletter

\providecommand{\tabularnewline}{\\}

\@ifundefined{textcolor}{}
{%
 \definecolor{BLACK}{gray}{0}
 \definecolor{WHITE}{gray}{1}
 \definecolor{RED}{rgb}{1,0,0}
 \definecolor{GREEN}{rgb}{0,1,0}
 \definecolor{BLUE}{rgb}{0,0,1}
 \definecolor{CYAN}{cmyk}{1,0,0,0}
 \definecolor{MAGENTA}{cmyk}{0,1,0,0}
 \definecolor{YELLOW}{cmyk}{0,0,1,0}
}

\usepackage{array}
\usepackage{babel}
\usepackage{eucal}

\makeatother

\usepackage{babel}
\begin{document}

\title{Hierarchy of modes in an interacting system}

\author{O. Tsyplyatyev}

\affiliation{School of Physics and Astronomy, University of Birmingham, Birmingham, B15 2TT, UK}

\author{A. J. Schofield}

\affiliation{School of Physics and Astronomy, University of Birmingham, Birmingham, B15 2TT, UK}

\author{Y. Jin}

\affiliation{Cavendish Laboratory, University of Cambridge, J J Thomson Avenue, Cambridge, CB3 0HE, UK}

\author{M. Moreno}

\affiliation{Cavendish Laboratory, University of Cambridge, J J Thomson Avenue, Cambridge, CB3 0HE, UK}

\author{W. K. Tan}

\affiliation{Cavendish Laboratory, University of Cambridge, J J Thomson Avenue, Cambridge, CB3 0HE, UK}

\author{C. J. B. Ford}

\affiliation{Cavendish Laboratory, University of Cambridge, J J Thomson Avenue, Cambridge, CB3 0HE, UK}

\author{J. P. Griffiths}

\affiliation{Cavendish Laboratory, University of Cambridge, J J Thomson Avenue, Cambridge, CB3 0HE, UK}

\author{I. Farrer}

\affiliation{Cavendish Laboratory, University of Cambridge, J J Thomson Avenue, Cambridge, CB3 0HE, UK}

\author{G. A. C. Jones}

\affiliation{Cavendish Laboratory, University of Cambridge, J J Thomson Avenue, Cambridge, CB3 0HE, UK}

\author{D. A. Ritchie}

\affiliation{Cavendish Laboratory, University of Cambridge, J J Thomson Avenue, Cambridge, CB3 0HE, UK}

\date{\today}
\begin{abstract}
Studying interacting fermions in 1D at high energy, we find a hierarchy in the spectral weights of the excitations theoretically and we observe evidence for second-level excitations experimentally. Diagonalising a model of fermions (without spin), we show that levels of the hierarchy are separated by powers of $\mathcal{R}^{2}/L^{2}$,
where $\mathcal{R}$ is a length-scale related to interactions and $L$
is the system length. The first-level (strongest) excitations
form a mode with parabolic dispersion, like that of a renormalised single
particle. The second-level excitations produce a singular power-law
line shape to the first-level mode and multiple power-laws at the
spectral edge. We measure momentum-resolved tunneling of
electrons (fermions with spin) from/to a wire formed within a GaAs
heterostructure, which shows parabolic dispersion of the first-level mode and well-resolved spin-charge separation at low energy with appreciable interaction strength.
We find structure resembling the second-level excitations, which dies away quite rapidly at high momentum.
\end{abstract}


\maketitle

Identifying patterns within the spectrum of interacting quantum systems
is notoriously hard and has often been limited to low-energy/low-momentum
excitations (such as the Fermi-liquid \cite{NozieresBook} or Luttinger-liquid \cite{GiamarchiBook} descriptions). In this Letter we demonstrate,
theoretically and supported by experiment, that a hierarchy of modes
can emerge in interacting 1D systems controlled by system
length. The dominant mode for long systems is a renormalized single-particle-like excitation which extends to high energies and momenta. This contrasts
with the well-known and distinctly non-free-particle-like description
of 1D systems at low energy. We show how these two regimes are connected.

\begin{figure}
\begin{centering}
\includegraphics[width=0.95\columnwidth]{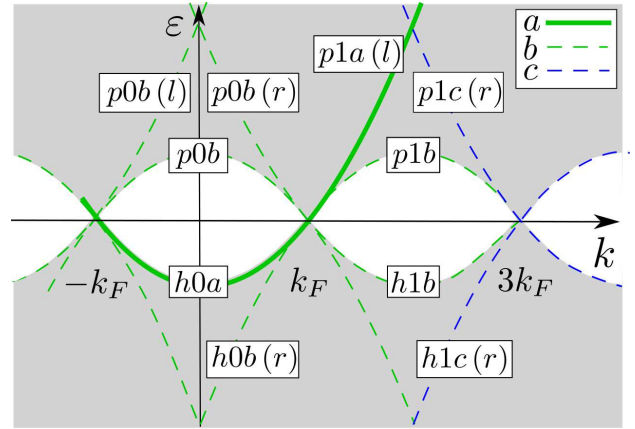} 
\par\end{centering}

\protect\caption{\label{fig:SF_spectral_function}The main features of spectral function for spinless fermions in the region $-k_{\rm F}<k<k_{\rm F}$ ($k_{\rm F}<k<3k_{\rm F}$) labelled by $0\left(1\right)$. The grey area marks non-zero values,
$p\left(h\right)$ shows the particle(hole) sector, $k_{\rm F}$ is the Fermi momentum, $a,\; b,\; c$ respectively identify the
level in the hierarchy in powers $0,\;1,\;2$ of $\mathcal{R}^{2}/L^{2}$,
and $\left(r,l\right)$ specifies the origin in the range---modes on the edge have no such label. }
\end{figure}

Our method is the diagonalization of a model of spinless fermions
with short-range interactions and the evaluation of its spectral function
via Bethe ansatz methods. We find that the spectral weights of excitations
have factors with different powers of a ratio of lengths, $\mathcal{R}^{2}/L^{2}$,
(which will be defined below) separating them into a hierarchy. The
dispersion of the mode formed by excitations with zero power (which
we call the first level%and label \textit{h0a} in Fig.\@ \ref{fig:SF_spectral_function}
) is parabolic with a mass renormalised by the Luttinger parameter $K$. The continuous spectrum of the second-level excitations produces a power-law line-shape around the first-level mode with a singular exponent $-1$. The particle edge \textit{p0b} in Fig.\@ \ref{fig:SF_spectral_function} 
% \highlitq{$k_{\rm F}$ is the Fermi wavevector}) 
is a second-level mode with a different power-law behaviour of the spectral function from that of the first-level mode \textit{h0a} of the hole edge, with an exponent that coincides with the mobile-impurity model \cite{TS13,GlazmanReview12}. The local
density of states is dominated by the first-level excitations, producing
a $1/\sqrt{\varepsilon}$ Van Hove singularity, where $\varepsilon$ is the energy measured the bottom of the conduction band, 
on the $\varepsilon>0$ side. For $\varepsilon<0$, the second-level excitations give the same $1/\sqrt{\left|\varepsilon\right|}$ singularity but with a different prefactor.

Experimentally, we measure momentum-resolved tunneling of electrons (fermions with spin) confined to a 1D geometry in the top layer of a GaAs-AlGaAs double-quantum-well structure from/to a two-dimensional electron gas in the bottom layer. We observe a single parabola (which particle-hole asymmetry also affects by increasing relaxation processes \cite{Yacoby10}) 
%in the 1D spectral function 
at high energy together with well-resolved spin-charge separation at low energy with appreciable interaction strength (ratio of charge and spin velocities $v_{\rm c}/v_{\rm s}\approx 1.4$) \cite{Yacoby02,Jompol09}. In addition, we can now resolve structure just above $k_{\rm F}$ that appears to be the edge of the second-level excitations (\textit{p1b}). However, for higher $k$ we find no sign of the higher-level excitations, implying that their amplitude must have become at least three orders of magnitude weaker than for the first parabola (\textit{h0a}).

\textit{Spinless fermions.} We study theoretically the model of interacting
Fermi particles without spin in 1D,
\begin{equation}
H=\int_{-\frac{L}{2}}^{^{\frac{L}{2}}}dx\left(-\frac{1}{2m}\psi^{\dagger}\left(x\right)\Delta\psi\left(x\right)-UL\rho\left(x\right)^{2}\right),\label{eq:H}
\end{equation}
where the field operators $\psi\left(x\right)$ satisfy the Fermi
commutation relations,$\left\{ \psi\left(x\right),\psi^{\dagger}\left(x'\right)\right\} =\delta\left(x-x'\right)$,
$\rho\left(x\right)=\psi^{\dagger}\left(x\right)\psi\left(x\right)$
is the particle density operator, and $m$ is the bare mass of a single
particle. Below we consider the periodic boundary condition, $\psi\left(x+L\right)=\psi\left(x\right)$,
restrict ourselves to repulsive interaction $U>0$ only, and take $\hbar=1$. 

In the Bethe ansatz approach the model in Eq.\@ (\ref{eq:H}) is diagonalised
by $N$-particle states parameterised with sets of $N$ quasimomenta
$k_{j}$ that satisfy the non-linear equations $\mathcal{L}k_{j}-2\sum_{l\neq j}\varphi_{jl}=2\pi I_{j}$
\cite{KorepinBook}, where $e^{i2\varphi_{ll'}}=-\left(e^{i\left(k_{l}+k_{l'}\right)}+1-2mUe^{ik_{l}}\right)/\allowbreak\left(e^{i\left(k_{l}+k_{l'}\right)}+1-2mUe^{ik_{l'}}\right)$
are the scattering phases and $I_{j}$ are sets of non-equal integer
numbers. The dimensionless length of the system $\mathcal{L}=L/\mathcal{R}$
is normalised by the short length-scale $\mathcal{R}$ which is introduced
using a lattice (with next-neighbor interaction) as the lattice parameter
(and interaction radius) $\mathcal{R}$ that provides microscopically
an ultraviolet cutoff for the theory. The latter procedure at high
energy is analogous to the point-splitting technique \cite{vonDelftSchoellerReview}
at low energy. For a small $\mathcal{R}$ the scattering phases become
linear functions of quasimomenta making the non-linear Bethe-ansatz
equations a linear system of equations \cite{TS14}. Solving them
for $\mathcal{L}\gg1$ via perturbation theory up to the first subleading
order in $1/\mathcal{L}$ we obtain

\begin{equation}
k_{j}=\frac{2\pi I_{j}}{\mathcal{L}-\frac{mUN}{mU+1}}-\frac{mU}{mU+1}\sum_{l\neq j}\frac{2\pi I_{l}}{\left(\mathcal{L}-\frac{mUN}{mU+1}\right)^{2}}.\label{eq:kj}
\end{equation}
The corresponding eigenenergy and total momentum (protected by the
translational invariance of the system) are $E=\sum_{j}k_{j}^{2}/\left(2m\right)$
and $P=\sum_{j}k_{j}$.

The response of a many-body system to a single-particle excitation
at momentum $k$ and energy $\varepsilon$ is described by a spectral
function that, in terms of the eigenstates, reads as \cite{AGD}
$A\left(k,\varepsilon\right)=\mathcal{L}\sum_{f}\Big[\left|\left\langle f|\psi^{\dagger}\left(0\right)|0\right\rangle \right|^{2}\allowbreak\delta\left(\varepsilon-E_{f}+E_{0}\right)\allowbreak\delta\left(k-P_{f}\right)+\left|\left\langle 0|\psi\left(0\right)|f\right\rangle \right|^{2}\allowbreak\delta\left(\varepsilon+E_{f}-E_{0}\right)\allowbreak\delta\left(k+P_{f}\right)\Big]$,
where $E_{0}$ is the energy of the ground state $\left|0\right\rangle $,
and $P_{f}$ and $E_{f}$ are the momenta and the eigenenergies of
the eigenstates $\left|f\right\rangle $; all eigenstates are assumed
normalised. 
\begin{table}
\begin{ruledtabular} %
\begin{tabular}{r|>{\centering\arraybackslash}m{2.75cm}|>{\centering\arraybackslash}m{3.75cm}}
$\textrm{ }$  & $x=0$  & $x=1$\tabularnewline
\hline 
$pxa$  & $-$  & $1$\tabularnewline
\hline 
$hxa$  & $1$  & $-$\tabularnewline
\hline 
$pxb$  & $\frac{16Z^{2}k_{\rm F}^{2}k^{2}}{\left(k^{2}-\left(k_{\rm F}+\gamma\right)^{2}\right)^{2}}$  & $\frac{4Z^{2}\gamma^{2}\left(k-k_{\rm F}+\frac{3}{2}\gamma\right)^{2}}{\left(k-k_{\rm F}+\gamma\right)^{2}\left(k-k_{\rm F}+2\gamma\right)^{2}}$\tabularnewline
\hline 
$pxb\left(l\right)$  & $\frac{4Z^{2}\left(k_{\rm F}+k\right)^{2}}{k_{\rm F}^{2}}$  & $-$\tabularnewline
\hline 
$pxb\left(r\right)$  & $\frac{4Z^{2}\left(k_{\rm F}-k\right)^{2}}{k_{\rm F}^{2}}$  & $-$\tabularnewline
\hline 
$hxb$  & $-$  & $\frac{4Z^{2}\left(3k_{\rm F}-k-\gamma\right)^{2}\left(k_{\rm F}+k\right)^{2}}{k_{\rm F}^{2}\left(k-k_{\rm F}+\gamma\right)^{2}}$\tabularnewline
\hline 
$hxb\left(l\right)$  & $\frac{4Z^{2}\gamma^{2}}{\left(k+k_{\rm F}+2\gamma\right)^{2}}$  & $\frac{Z^{2}k_{\rm F}^{2}k^{2}}{\left(\left(k+\gamma\right)^{2}-k_{\rm F}^{2}\right)^{2}}$\tabularnewline
\hline 
$hxb\left(r\right)$  & $\frac{4Z^{2}\gamma^{2}}{\left(k-k_{\rm F}-2\gamma\right)^{2}}$  & $-$\tabularnewline
\end{tabular}\end{ruledtabular}\protect\caption{\label{tab:SF_spectral_function_values}Spectral weights along the
$a$- and the $b$-modes for $-k_{\rm F}<k<k_{\rm F}$($k_{\rm F}<k<3k_{\rm F}$)
labeled by $x=0\left(1\right)$. Terminology is the same as in Fig.\@ \ref{fig:SF_spectral_function} and $\gamma=2\pi/\mathcal{L}$.
}
\end{table}
We now turn to calculation of the form factors using the algebraic
form of the Bethe ansatz \cite{AGD}. To proceed we borrow results from
Heisenberg chains: the normalisation of Bethe states was obtained
in Ref.\@ \onlinecite{Gaudin1981} and the matrix elements were obtained \cite{Kitaine99_00}
using Slavnov's formula for scalar products \cite{Slavnov89} and
Drinfield twists \cite{Drinfeld86} to represent the field operators
in the basis of Bethe states. The anti-commutativity of Fermi particles
at different positions can be easily added by introducing a fermionic
basis for auxiliary space in the construction of the algebraic Bethe ansatz
 \cite{Korepin00Sakai08}, which results in a determinant expression
for $\left\langle f|\psi^{\dagger}\left(0\right)|0\right\rangle $.
For a small $\mathcal{R}$, which corresponds to low particle density
we obtain \cite{Eq3check} 
\begin{multline}
\left|\left\langle f|\psi^{\dagger}\left(0\right)|0\right\rangle \right|^{2}=\frac{Z^{2N}}{\mathcal{L}}\frac{\prod_{j}\left(k_{j}^{0}-P_{f}\right)^{2}}{\prod_{i,j}\left(k_{j}^{f}-k_{i}^{0}\right)^{2}}\\
\prod_{i<j}\left(k_{j}^{0}-k_{i}^{0}\right)^{2}\prod_{i<j}\left(k_{j}^{f}-k_{i}^{f}\right)^{2},\label{eq:FF_N}
\end{multline}
where $Z=mU/\left(mU+1\right)/\left(\mathcal{L}-NmU/\left(1+mU\right)\right)$
and $k_{j}^{f}$ and $k_{j}^{0}$ are quasimomenta of the eigenstate
$\left|f\right\rangle $ and the ground state $\left|0\right\rangle $. 

This result is singular when one or more quasimomenta of an excited
state coincide with that of the ground state. The divergences occur
in the leading order of Eq.\@ (\ref{eq:kj}) but the first subleading
order provides a cutoff within the theory cancelling a power of $Z^{2}\sim\mathcal{L}^{-2}$
per singularity; when $N$ quasimomenta $k_{j}^{f}$ coincide with
$k_{j}^{0}$, Eq.(\ref{eq:FF_N}) gives $\mathcal{L}\left|\left\langle f|\psi^{\dagger}\left(0\right)|0\right\rangle \right|^{2}=1$.
We label the many-body excitations by the remaining powers of $\mathcal{L}^{-2}$
\cite{Hierarchy}, e.g. $p0b$: $p\left(h\right)$ indicates the particle
(hole) sector, $0\left(1\right)$ encodes the range of momenta $-k_{\rm F}<k<k_{\rm F}$ $\left(k_{\rm F}<k<3k_{\rm F}\right)$,
and $a,\; b,\; c$ reflect the terms $\mathcal{L}^{-2n}$ with
$n=0,\;1,\;2$. All simple modes, formed by single particle- and hole-like
excitations of the ground state $k_{j}^{0}$, are presented in Fig.
\ref{fig:SF_spectral_function} and the spectral function along them
is evaluated in Table \ref{tab:SF_spectral_function_values}. Note
that the thermodynamic limit involves both ${\cal L\rightarrow\infty}$
and the particle number $N\rightarrow\infty$ and the finite combination
$N/{\cal L}$ ensures that the spectral weight of sub-leading modes,
e.g. the modes $p0b$, $h1b$, and $h1b\left(r\right)$, is still
apparent in the infinite system.

Excitations around the strongest $a$-modes have an additional electron-hole
pair in their quasimomenta, which introduces an extra factor of $\mathcal{L}^{-2}$, 
\begin{equation}
\left|\left\langle f|\psi^{\dagger}\left(0\right)|0\right\rangle \right|^{2}=\frac{Z^{2}}{\mathcal{L}}\frac{\left(k_{2}^{f}-k_{1}^{f}\right)^{2}\left(k_{1}^{0}-P_{f}\right)^{2}}{\left(k_{1}^{f}-k_{1}^{0}\right)^{2}\left(k_{2}^{f}-k_{1}^{0}\right)^{2}}.\label{eq:FF_2}
\end{equation}
The energies of the electron-hole pairs themselves are regularly spaced
around the Fermi energy with slope $v_{\rm F}$. However, degeneracy
of the many-body excitations due to the spectral linearity makes the
level spacings non-equidistant. Using a version of the spectral function smoothed over energy, $\overline{A}\left(\varepsilon\right)=\int_{-\epsilon_{0}/2}^{\epsilon_{0}/2}d\epsilon A\left(\varepsilon+\epsilon,k\right)/\epsilon_{0}$
where $\epsilon_{0}$ is a small energy scale, we obtain  $\overline{A}\left(\varepsilon\right)=Z^{2}\allowbreak2k_{\rm F}\left(3k^{2}+k_{\rm F}^{2}\right)\allowbreak/\left(m\gamma K\right)\allowbreak\left(\varepsilon_{h0a}-\varepsilon\right)^{-1}\allowbreak\theta\left(\varepsilon_{h0a}-\varepsilon\right)$
and $\overline{A}\left(\varepsilon\right)=Z^{2}\allowbreak\left(k+\textrm{sgn}\left(\varepsilon-\varepsilon_{p1a\left(l\right)}\right)k_{\rm F}\right)^{3}\allowbreak/\left(m\gamma K\right)\allowbreak\left|\varepsilon-\varepsilon_{p1a\left(l\right)}\right|^{-1}$
\cite{a_validity}, where $\gamma=2\pi/\mathcal{L}$ and the dispersion of the $a$-modes is parabolic $\varepsilon_{h0a}\left(k\right)=\varepsilon_{p1a\left(l\right)}\left(k\right)=k^{2}/\left(mK\right)$
with the mass renormalised by the Luttinger parameter $K$, around the $h0a$ and $p1a\left(l\right)$ modes. The exponent
$-1$ coincides with the prediction of the mobile impurity model \cite{TS13}
where the spectral edge is an $a$-modes, $h0a$. 

Excitations around $b$-modes belong to the same level of hierarchy
as the modes themselves, Eq.\@ (\ref{eq:FF_2}), giving a more complicated
shape of the spectral function. Let us focus on one mode, $p0b$. It
has a new power-law behaviour characterised by an exponent changing
with $k$ from $\overline{A}\left(\varepsilon\right)\sim\left(\varepsilon-\varepsilon_{p0b}\right)^{3}$
for $k=0$ to $\overline{A}\left(\varepsilon\right)\sim\textrm{const}-\left(\varepsilon-\varepsilon_{p0b}\right)$
for $k\approx k_{\rm F}$, where $\varepsilon_{p0b}\left(k\right)=k_\textrm{F}^2/\left(mK\right)-k^{2}/\left(mK\right)$. 
This is essentially different from predictions of the mobile-impurity
model. Here we observe that the phenomenological model in Refs.\@ \onlinecite{Khodas06Imambekov09}
is correct only for the $a$-mode spectral edge but higher-order edges
require a different field-theoretical description. The density of
states is linear, $\nu\left(\varepsilon\right)\sim\left(\varepsilon-\varepsilon_{p0b}\right)$,
but level statistics varies from having a regular level spacing (for
$k$ commensurate with $k_{\rm F}$) to an irregular distribution (for
incommensurate $k$), which is another microscopic difference between
$a$- and $b$-modes.

\begin{figure}
\centering{}\includegraphics[width=0.95\columnwidth]{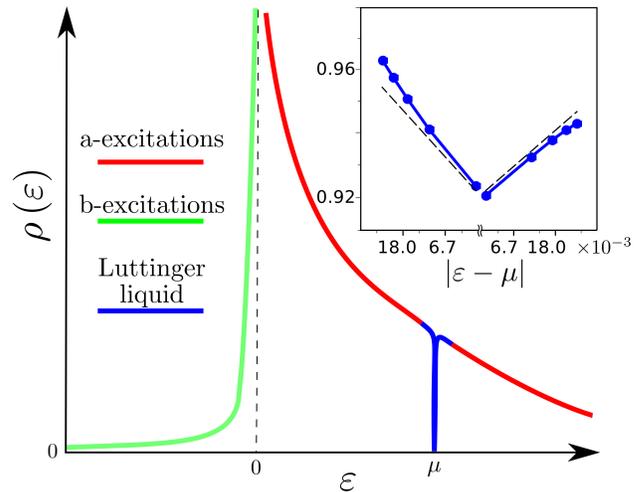}\protect\caption{\label{fig:SF_LDOS}The local density of states for spinless fermions:
red and green lines show the contribution of $a$- and $b$-excitations
and the blue line indicates the Luttinger-liquid regime. Inset is a log-log
plot around the Fermi energy: the blue points are numerical data for
$N=71$, $L=700$, $mV=6$ giving $K=0.843$, and the dashed line is $\rho\left(\varepsilon\right)=\textrm{const}\left|\varepsilon-\mu\right|^{\left(K+K^{-1}\right)/2-1}$.}
\end{figure}
Now we use the result in Eq.\@ (\ref{eq:FF_N}) to calculate another
observable, the local density of states. It is independent of position
for translationally invariant systems and, in term of eigenmodes,
is \cite{AGD,rho_GF} $\rho\left(\varepsilon\right)=\mathcal{L}\sum_{f}\Big[\left|\left\langle f|\psi^{\dagger}\left(0\right)|0\right\rangle \right|^{2}\allowbreak\delta\left(\varepsilon-E_{f}+E_{0}\right)\allowbreak+\left|\left\langle 0|\psi\left(0\right)|f\right\rangle \right|^{2}\allowbreak\delta\left(\varepsilon+E_{f}-E_{0}\right)\allowbreak\Big]$.
The leading contribution for $\varepsilon>0$ comes from $a$-modes,
$\rho\left(\varepsilon\right)=\theta\left(\varepsilon\right)\sqrt{2mK/\varepsilon}$,
which gives the same $1/\sqrt{\varepsilon}$ functional dependence
as the free-particle model---see red line in Fig.\@ \ref{fig:SF_LDOS}. Around the Fermi
energy the Tomonaga-Luttinger model predicts power-law suppression
of $\rho\left(\varepsilon\right)\sim\left|\varepsilon-\mu\right|^{\left(K+K^{-1}\right)/2-1}$
\cite{GiamarchiBook} (blue region in Fig.\@ \ref{fig:SF_LDOS}) signaling that the
leading-order expansion in the $\mathcal{L}\left|\left\langle f|\psi^{\dagger}\left(0\right)|0\right\rangle \right|^{2}=1$
result is insufficient. We evaluate $\rho\left(\varepsilon\right)$
numerically in this region using  determinant representation of the
form factors for the lattice model instead of Eq.\@ (\ref{eq:FF_N}) (inset in Fig.\@ \ref{fig:SF_LDOS}) \cite{HELLoverlap}. Away from
the point $\varepsilon=\mu$ the particle-hole symmetry of the Tomonaga-Luttinger
model is broken by the curvature of the dispersion of the $a$-modes. For $\varepsilon<0$
the leading contribution to $\rho\left(\varepsilon\right)$ comes
from $b$-modes. Using Eq.\@ (\ref{eq:FF_2}) we obtain $\rho\left(\varepsilon\right)=2Z^{2}k_{\rm F}^{2}/\left(\gamma\mu K\right)\big[2\left(1-3\left|\varepsilon\right|/\mu\right)\sqrt{\mu}\allowbreak\cot^{-1}\left(\sqrt{\left|\varepsilon\right|/\mu}\right)\allowbreak/\sqrt{\left|\varepsilon\right|}+6\big]\theta\left(-\varepsilon\right)$,
which contains another Van Hove singularity $\rho\left(\varepsilon\right)=2\pi Z^{2}k_{\rm F}^{2}/\left(\gamma K\sqrt{\mu\varepsilon}\right)$
at the bottom the conduction band (green line in Fig.\@ \ref{fig:SF_LDOS}).

\begin{figure}
\includegraphics[width=1\columnwidth]{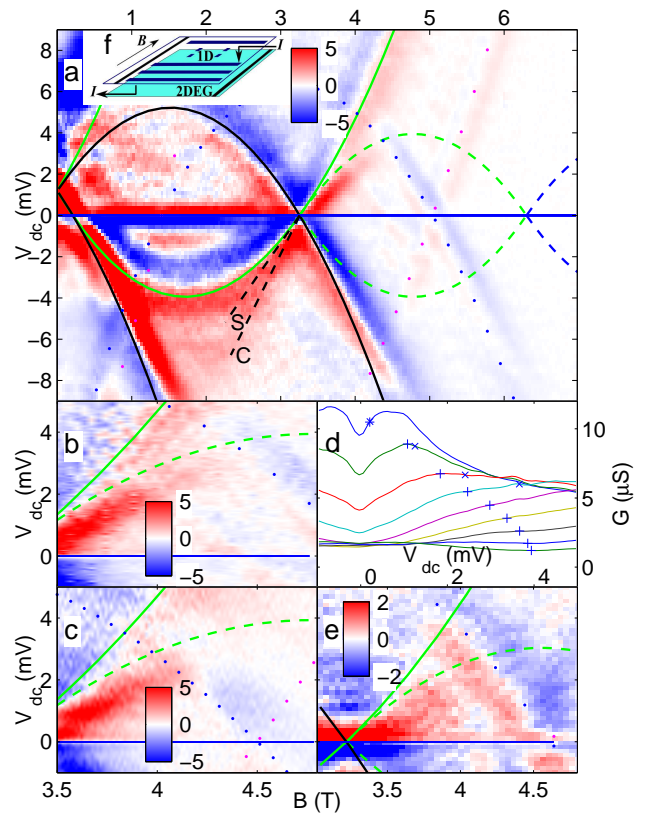}\protect\caption{\label{fig:experiment_main} Measurement of the tunneling differential conductance $G={\rm d}I/{\rm d}V$
for two samples, each consisting of a set of identical wires of length $L=10\mu$m  (a-d) and $L=18\mu$m (e), sketched in inset f.
(a) Intensity plot of ${\rm d}G/{\rm d}V_{\rm dc}$ in the range $(-k_{\rm F}..3k_{\rm F},-2\mu..2\mu)$. The green solid line marks $a$-modes,
dashed green lines, $b$-modes and dashed blue, $c$-modes (as in Fig.\@ \ref{fig:SF_spectral_function}); dotted magenta and blue lines are parasitic 2D dispersions. Spin (S) and charge (C) modes are indicated with black dashed lines. $T=300$\,mK.
(b) Enlargement of the replica feature in (a) just above $k_{\rm F}$. It appears as a pale band (slowly varying $G$) between the two green curves, after a red region (sharp rise in $G$). (c) The same as in (b), but with the gate voltage over most of the parasitic (`p') region changed to move the `p' parabolae.
(d) $G$ $vs$ $V_{\rm dc}$ at various fields $B$ from 3 to 4.8\,T (from (a)); `+' and `$\times$' symbols on each curve indicate, respectively, the voltages corresponding to the dashed and solid ($p1b$ and $p1a(l)$) green lines in (a) and (b), showing the enhanced conductance between the two.
(e) ${\rm d}G/{\rm d}V_{\rm dc}$ for a second device, at $T<100$\,mK. The replica feature is similar to that shown in (b) and (c) for the other sample.
}
\end{figure}

\textit{Fermions with spin.} We study experimentally spin-unpolarised
electrons in a high-mobility GaAs-AlGaAs double-quantum-well
structure with electron density around $2 \times 10^{15}$\,m$^{-2}$ in each layer. Electrons in the top layer are confined to a 1D geometry
by split gates. Our devices contain an array of $\sim$$500$ highly regular
wires to boost the signal from 1D-2D tunneling. The small lithographic
width of the wires, $\sim$$0.18$\,$\mu$m, provides a large energy spacing between the first and second 1D subbands, allowing a wide energy window for electronic excitations in the single-subband case---see device schematic in Fig.\@ \ref{fig:experiment_main}f and more details in Ref.\@ \onlinecite{Jompol09}. 

The 2DEG in the bottom layer is separated from the wires by a $d=14$\,nm tunnel barrier (giving a spacing between the centres of the wavefunctions of $d=36$\,nm). It
is used as a controllable injector or collector of electrons for the
1D system \cite{Altland99}. A sharp spectral feature in the density
of states of the 2DEG produced by integration over momenta in the direction
perpendicular to the wires can be shifted in energy by a dc-bias between the layers, in order to probe different energies. Also, an in-plane magnetic field
$B$ applied perpendicular to the wires changes the longitudinal
momentum in the tunneling between layers by $\Delta k=eBd/\hbar$,
where $e$ is the electronic charge, and so probes the momentum. Together they reveal the dispersion relation of states in each layer. In this magnetic field range the system is still within the regime of Pauli paramagnetism for the electron densities in our samples.

We have measured the tunneling conductance $G$ between the two layers (see Fig.\@ \ref{fig:experiment_main}f) in detail in a wide range of voltage and magnetic field, corresponding to a large portion of the 1D spectral function from $-k_{\rm F}$ to $3k_{\rm F}$ and from $-2\mu$ to $2\mu$ (Fig.\@ \ref{fig:experiment_main}a). At low energy we observe spin-charge separation \cite{Jompol09}. The slopes of the charge (C) and spin (S) branches---black dashed lines---are $v_{\rm c}\approx 2.03\times 10^5$ms$^{-1}$ and $v_{\rm s}\approx 1.44\times 10^5$ms$^{-1}$, respectively, with $v_{\rm c}/v_{\rm s}\approx 1.4 \pm 0.1$. This large ratio, together with a strong zero-bias suppression of tunneling \cite{Jompol09}, confirms that our system is in the strongly interacting regime.

Unavoidable `parasitic' (`p') tunneling from narrow 2D regions connecting the wires to the injector constriction \cite{Jompol09}, superimpose a set of parabolic dispersions, marked by magenta and blue dotted lines in Fig.\ref{fig:experiment_main}a \cite{capacitive_corrections}, on top of the 1D-2D signal. Apart from them we observe a single 1D parabola, marked by the solid green line in Fig.\@ \ref{fig:experiment_main}a, which extends from the spin-excitation branch at low energy. The position of its minimum gives the 1D chemical potential $\mu\approx 3$\,meV and its crossings with the line $V_{\rm dc}=0$, corresponding to momenta $-k_{\rm F}$ and $k_{\rm F}$, give the 1D Fermi momentum $k_{\rm F}\approx 8\times 10^7$\,m$^{-1}$.

All other edges of the 1D spectral function are constructed by mirroring and translation of the hole part of the observable 1D dispersion, dashed green and blue lines in Figs.\@ \ref{fig:experiment_main}. We observe a distinctive feature in the region just above the higher $V_{\rm dc}=0$ crossing point ($k_{\rm F}$): the 1D parabola, instead of just continuing along the non-interacting parabola, broadens, with one boundary following the parabola ($p1a(l)$) and the other bending around, analogous to the replica $p1b$. This is observed in samples with different wire designs and lengths (10\,$\mu$m (a-d), and 18\,$\mu$m, (e)) and at temperatures from 100\,mK up to at least 300\,mK. The strength of the $p1b$ feature decreases as the $B$ field increases away from the crossing point analogously to that for spinless fermions in Table\@ \ref{tab:SF_spectral_function_values}, though it then passes a `p' parabola. (b) and (c) show the replica feature for two different positions of the `p' parabolae using a gate above most of the `p' region, showing that the replica feature is independent of the `p' tunneling. $G$ is plotted in (d) on cuts along the $V_{\rm dc}$ axis of (a) at various fields $B$ from 3 to 4.8\,T; between the `+' and `$\times$' symbols on each curve is the region of enhanced conductance characteristic of the replica $p1b$. The amplitude of the feature dies away rapidly, and beyond the `p' parabolae, we have measured up to 8T with high sensitivity, and find no measurable sign of any feature above the experimental noise threshold. This places an upper limit on the amplitude of any replica away from $k_{\rm F}$ of at least three orders of magnitude less than that of the $a$-mode ($h0a$).

Making an analogy with the microscopic theory for spinless fermions
above, we estimate the ratio of signals around different spectral
edges using the 1D Fermi wavelength, $\lambda_{\rm F}\approx 80$\,nm for our
samples, as the short-range scale.
The amplitude of signal from the second (third)-level excitations
is predicted to be smaller by a factor of more than $\lambda_{\rm F}^{2}/L^{2}=6\times 10^{-5}$
($\lambda_{\rm F}^{4}/L^{4}=4\times 10^{-9}$), where the length
of a wire is $L=10\,\mu$m. These values are at least an order of
magnitude smaller than the noise level of our experiment. Thus, our
observations are consistent with the mode hierarchy picture for fermions.

\emph{In conclusion}, we have shown that a hierarchy of modes can emerge in an interacting 1D system controlled by the system length. The dominant mode for long systems has a parabolic dispersion, like that of a renormalised free particle, in contrast with distinctly non-free-particle-like behaviour at low energy governed by the Tomonaga-Luttinger model. Experimentally we find a clear feature resembling the second-level excitations, which dies away at high momentum.

We acknowledge financial support from the UK EPSRC through Grant No.\@ EP/J01690X/1 and EP/J016888/1.


\begin{thebibliography}{10}
\bibitem{Eq3check} The result in Eq.\@ (\ref{eq:FF_N}) was verified by direct summation over spatial variables for up to $N=3$ particles and details of its derivation will be published elsewhere.
\bibitem{a_validity} The exponent $-1$ is valid while the level
spacings remain regular, $\left|\varepsilon-\varepsilon_{p1a\left(h0a\right)}\right|\apprle k_{\rm F}^{2}/\left(mK\right)$.
Below the $p1a$ mode, the spectral edge gives an additional restriction,
$\varepsilon_{p1a}-\varepsilon\apprle\left(k_{\rm F}-k\right)^{2}/\left(mK\right)$.
\bibitem{rho_GF}It is equivalent to conventional definition, $\rho\left(x,\varepsilon\right)=-\textrm{Im}\big[\int dte^{-i\varepsilon t}G(x,x,t)\big]\textrm{sgn}(\varepsilon-\mu)/\pi$, using the two point correlation
function $G\left(x,x',t\right)=-i\left\langle T\left(e^{-iHt}\psi\left(x\right)e^{iHt}\psi^{\dagger}\left(x'\right)\right)\right\rangle $
at zero temperature.
\bibitem{HELLoverlap}The regions of validity of the Tomonaga-Luttinger
model and Eq.\@ (\ref{eq:FF_N}) have a large overlap around $\varepsilon=\mu$
since power-law suppression of $\rho\left(\varepsilon\right)$ has
a very small exponent due to only small deviations from $K=1$ for
arbitrary short-range interactions between spinless fermions.
\bibitem{capacitive_corrections}All the parabolae have been modified to attempt to take into account the capacitance between the layers, see details in Ref.\@ \onlinecite{Jompol09}.
\end{thebibliography}
\end{document}